 \pdfoutput=1
\documentclass{svjour2}                    
\smartqed  
\usepackage{graphicx}
\journalname{arxiv}
\begin{document}

\title{Symmetry in Shannon's Noiseless Coding Theorem  
}


\author{L. F. Johnson       
}


\institute{L. F. Johnson \at
              University of Waterloo  \\
              \email{lfjsde@gmail.com}           
}

\date { 2010/10/29}

\maketitle
\begin{abstract}
Statements of Shannon's Noiseless Coding Theorem by various authors, 
including the original, are reviewed and clarified. Traditional statements of the 
theorem  are often unclear as to when it applies. A new notation is introduced 
and the domain of application  is clarified.

An examination of the bounds of  the Theorem 
leads to a new symmetric restatement. It is shown that the extended upper bound 
is an acheivable upper bound, giving symmetry to the theorem.

The relation of information entropy to the physical entropy of Gibbs and Boltmann is illustrated. 
Consequently, the study of Shannon Entropy is strongly related to physics and there is a physical theory of information. This paper
is the beginning of of an attempt to clarify these relationships.

\keywords{Shannon coding \and information theory \and entropy \and coding compression 
\and information metrics  \and thermodynamics }
\end{abstract}

\section{Introduction}

An important result of Shannon [1948] shows that it is possible 
to approach 100\% coding efficiency, when no noise is present [1].
This result has various names: Shannon's noiseless coding theorem, 
Shannon's first theorem, the Data compression theorem, and oddly enought is 
often not even stated as a theorem but only as a formula. 
This paper refers to it as the Theorem.

Several definitions are needed to make our notation clear.
A set of events $S$ is indexed by $i$, and the event $i$ is represented 
by the symbol $s_{i}$, which  has probability $p_{i}$. There are two sets of 
symbols: the input alphabet $S$, of size q, and the code alphabet $A$, 
of size $r$.
 The length of the code for event $i$
is $l_{i}$. The expected code length of a code $C$ 
on input symbols $S$ is $\overline{L}_{C}(S)$.

\[ H_{r}(S)=\sum_{S} p_{i} \log_{r} \frac{1}{p_{i}} \]
\[ \overline{L}_{C}(S) =\sum_{i=1}^{q} p_{i} l_{i}  \]

We call the entropy to the base r the {\em r-entropy}. A code of 
radix r is an {\em r-code}.
 The entropy of a 
set $S$ in bits is $H(S)$.
\[ H(S)=\sum_{S} p_{i} \lg \frac{1}{p_{i}} \]

\subsection{Physical Entropy}
Shannon Entropy $H$ and Gibbs Entropy ${\cal S}$ are abstractly related.
In a classical system with a discrete set of micro states S where $E_i$
is the energy of a microstate with probability $p_i$,  then the physical Gibbs Entropy is: 

\[ {\cal S}(S)= k_B\sum_{S} p_{i} \ln  \frac{1}{p_{i}}  \ \ J/K\]  
Where $k_B$ is the Boltzmann constant $k = 1.380?6504(24)   \times 10^{-23}$  J/K.

This reduces to information entropy by algabracic transform.

\[ H = {\cal S} (S)/(k_B \ln 2)=  \sum_{S} p_{i} \lg \frac{1}{p_{i}} \  bits\]

Boltzmann Entropy

\[ {\cal S } = k_B \ln {\Omega } \]

$\Omega$ is the number of microstates consistent with the given macrostate. This is the equilikely case of maximum 
entropy.

\[ H =  {\cal S }/(k_B  \ln 2)  = \lg {\Omega }   = \lg n \]

The statistical entropy reduces to Boltzmann's entropy when all the 
accessible microstates of the system are equally likely. It is the 
configuration corresponding to the maximum of a system's entropy for a 
given set of accessible microstates. Here 
 the lack of information is maximal. As such,  Boltzmann's entropy is the 
expression of entropy at thermodynamic equilibrium in the micro-canonical
 ensemble. 
Consequentally, the study of Shannon Entropy is strongly related to physics [2].
Naturally, the number of information states is much less than in physical entropy except when Quantum 
Information is studied but that is a future paper.

\section{History}
In this section, the notation of each of the orginal authors will be used, when 
possible, and some care is needed by the reader. 
To reduce confusion, we will normalize the terminology where necessary 
in the representations by various authors. Hopefully, the spirit of 
their representations will show even if not quite always literally faithful.

Essentially, the Theorem says that we can construct a code with an 
average number of
code symbols per input event that is as close to the entropy of $S$ as we like. 
First, we look at how the Theorem is stated by a number of authors.
A typical statement of the Theorem is given by Hamming. 

{\bf The Theorem} (from Hamming[2] 1986)
$ H_{r}(S) \leq L < H_{r}(S) +\frac{1}{n} $

The code has r symbols and $L$ is the average code word length per input 
symbol or our $\overline{L}$.

{\bf The Theorem} (from Abramson[3] 1963)
\[ H_{r}(S) \leq \frac{L_{n}}{n} < H_{r}(S) +\frac{1}{n} \]

Here the use of the nth extension coding is indicated.

{\bf The Theorem} (from Gallager[4] 1968)
... it is possible to assign code words ...
\[ \frac{H(U)}{\log D} \leq \overline n < \frac{H(U)}{\log D} +\frac{1}{L} \]

Since the essence of the theorem is that $\overline{L}$  can approach the
entropy, the Theorem can also be expressed as a limit.

{\bf The Theorem} (from McEliece[5] 1977)
$ \lim_{m \rightarrow \infty} \frac{n_{s}(p^{m})}{m}=H_{s}(p)$

Here the connection that the entropy is measured in the radix of the 
code is indicated.
 

{\bf The Theorem} (from Cover[6] 1991)
$ H(S) \leq L_{n}^{\ast} < H(S) + \frac{1}{n}$

This assumes binary coding. $L_{n}^{\ast}$ is the average code word 
length per input symbol of an optimum code. Here a restriction on 
the form of coding is indicated and the Theorem restricted to 
optimum codes.

{\em A. Shannon's  Original Statement}

If we return to Shannon's original work, we see a more complex view and 
one that makes it more difficult to derive. 
Shannon first expressed the theorem in terms of 
channel capacity; however, he also gave the essence of the theorem in 
terms of entropy in an alternate proof that leads to all the 
statements in this paper. 
To begin with, Shannon used 
an estimator for  the entropy of the source. He considered all 
sequences $S_{i}$ of $N$ symbols in the source, thus, determining $H$ from the 
statistics of the message sequence.

{\bf The Theorem} (after Shannon 1948)
\[ G_{N} = \frac{1}{N} \sum_{i} p(S_{i}) \lg \frac{1}{p(S_{i})} \] 
\[ \lim_{N \rightarrow \infty }  G_{N} = H \]
\[  G_{N} \leq \overline{B} < G_{N} + \frac{1}{N} \]

Where $G_{N}$ approaches the entropy as $N$ increases, and $\overline{B}$ 
is the average number of binary symbols per symbol of the source.

Except for Gallager, it is usually not clear fom the  Theorem 
statement that this is only true for some codes. 
One must read the proofs to see that Shannon codes always satisfy 
the relation. While it is true for Shannon codes or better, it 
is also true of other codes, a fact generally not apparent from the 
discussion. The Theorem is important enough to have a nice 
complete compact statement.

\section{Analysis}


We now examine various ways of restating the Theorem.
When no code radix $r$ is specified, entropy will be measured in bits, 
and all logs are to the base 2, written as {\em lg}.
It is important to note that the entropy of the Theorem is measured in the 
radix of the code. Of course, \( H_{r}= H/ \lg r \).  

{\em A. The Code Bounding Lemma}

The lower bound on the average code length of a code is the 
r-entropy of the source set. This interprets entropy as the 
best possible expected code length for an r-code. 

\[ H_{r}(S) \leq \overline{L} \]

Discrete codes cannot always achieve this bound. 
For a uniform distribution, 
the average length of codes with $q=r^{k}$, where $k$ is an 
integer greater than zero, is equal to the entropy.
 
The surprising fact is that we can always find a code  better than the 
entropy plus one.
The usual upper bound proof for this assumes Shannon 
coding $\cal S$ or better. For simplicity, 
this is often expressed as the  bound for 
Shannon coding.

\[  \overline{L}_{\cal S} < H_{r}(S)+1 \]

When the radix of the code is binary, it is usual to drop the 
subscript $r$. Assuming binary simplifies notation in proofs, but 
obscures the fact that the entropy of these relations must be 
measured in the radix of the code used. 

{\em B. Correcting the Upper Bound}

Consider the case $H=0$
\[ 0 \leq   \overline{L}_{\cal S} < 0 + 1 \]

An entropy of 
zero is given by any distribution where exactly one event 
is certain. What is the value of $\overline L$ when $H=0$?
Both Shannon and Huffman codes will assign a 
code of length 0 to the certain event. The average length 
of the resulting code will be 0 in this special case.
Thus we have: $ 0 \leq  0 < 0 + 1 $

However, the Theorem is not restricted to these codes. 
Define an extended Shannon code as one that assigns 1 
to the certain event. The fact that this is less efficient is not important.
Thus we have: $ 0 \leq  1 < 0 + 1 $

This is impossible, so the upper bound must be achievable in this 
special case. 

Allowing a probability of zero results in infinite length 
code words. A simple further modification of the definition of a Shannon code 
gives finite codes. If $p_{i}=0$, then $l_{i}=0$. 
Our extended Shannon code is denoted by {\sf s}. 

{\em C. A New Upper Bound}

\[  \overline{L}_{{\sf s}} \leq H_{r}(S)+1 \]

There exist bounded codes that achieve the upper bound.
Why was  this result not observed before this? Certainly, a great many 
people have seen this relation.
The standard proof, for binary codes, begins with:

\[ \lg (\frac{1}{p_{i}}) \leq l_{i} < \lg (\frac{1}{p_{i}}) + 1 \]

because for a Shannon code \( l_{i}=\lceil \lg(1/p_{i}) \).

At this point to allow $p_{i}=0$ is obviously silly and so the 
boundary case of a certain event is never considered. Proofs are 
a demonstration of understanding, not a method of discovery.

Our result on the new upper bound for 
$\overline{L}_{C} \leq \overline{L}_{{\sf s}}$ is simple 
but apparently not obvious [1],[2],[3],[4],[5],[6]. It resulted from 
asking why the Theorem did not exhibit symmetry.

Should this special case be included? The  argument that 
can be used in the certain event boundary case 
is that when an event is certain, no code word 
need be used in any coding method, thus the average length 
is zero, and the original statement is correct. However in general 
for n codes, we do not know their exact probabilities and these may or 
may not be zero. In fact we would have to exclude the assignment 
of a code to the certain event. While using a Shannon code for 
the proof of the Theorem does in fact do this, codes less efficient 
than Shannon still satisfy the original theorem.   
We can conclude that the restatement of the upper bound 
is correct.

A final objection is that the chance of one event having a 
probability of one is unlikely and trivial.
This may be so, but equilikely events are even less probable as 
the number of events increases. 

{\em D. General Code Bounding Lemma}

There is no upper bound in general for $\overline{L}_{C}$, but we 
can define a working upper bound. Any S can be encoded 
by a block code B and $\overline{L}_{B}= \lceil \log_{r}q$.
The only reason to use a variable length code is to be more 
efficient than a block code. So it is reasonable to 
never use codes with average length worse than a block code. 
Define a good code, gC, as any code where 
$\overline{L}_{gC} \leq \overline{L}_{B}$ 
This gives a general form of the Lemma.

{\bf Lemma: Good Code Bounding}
\( H_{r} \leq \overline{L}_{gC} \leq H_{r}+ k \)

Where $k= \log_{r} q$, or k=1 for 
all \(\overline{L}_{C} \leq \overline{L}_{{\sf s}} \)

Actually, we have \( \overline{L}_{gC} \leq \lceil \log_{r} q \); 
thus, the lemma bound is not always good.

{\em E. The n th Extension of a Code}

\medskip
The Theorem is a consequence of recoding a source in blocks of source symbols 
in order to better match the code lengths to code probabilities.
The nth extension of a source is formed by concatenating all sequences 
of the original source symbols to form {\em a new (compound) symbol} or 
event. This new event now has a probability that is the 
product of the probabilities of the original source symbols. 

The new alphabet is $S^{n}$, and the average 
length of the extended code in blocks is  $\overline{L}(S^{n})$. 
The average length of the extended code in the input 
symbols $S$ is $\overline{L}(S)$.

{\bf Lemma: Extension Average Length}
\( \overline{L}(S^{n})=n \overline{L}(S) \)

When it is necessary to remember that the nth extension was encoded 
for $\overline L(S)$, 
we use the power $n$, as in $\overline L^{\ n}(S)$.
 
Our new definition, thus, extends nicely and is related to the 
entropy relation for code extensions.

{\bf Lemma: Extension Entropy}
  \[H(S^{n})= n H(S)\]


\section{Results}

We  now summarize the results of our discussion and 
obtain a new statement of the Theorem.

{\bf Theorem 1: Code Bounding}
For extended Shannon codes ${\sf s}$
\( H_{r}(S) \leq \overline{L}_{\sf s} \leq H_{r}(S)+1 \)

Further symmetries can be observed.

If $H = \overline L$, then $H = \overline L < H+1$.

If $H = 0 $, then $H < \overline L = H+1$.

The maximum excess of 1 is not always attained. For $q=r^{k}$ 
equally likely events, the excess is 
zero  but jumps to $(q-1)/q$ for {\em almost} equally 
likely events, where all but one are just over probability 
1/$q$. Shannon code excess can be quite sensitive to small changes 
in probability.

The new result follows from the known technique of coding 
the n th extension and the preceeding relations. 
We have renamed it the Code Compression Theorem of Shannon, to 
reflect what it says.

{\bf Theorem 2: Code Compression} For extended Shannon codes or better
\( H_{r}(S) \leq \overline{L}_{{\sf s}}^{\ n}(S) \leq  H_{r}(S)+\frac{1}{n} \)

{\bf Proof}:
Encode the nth extension of a source by an 
extended Shannon code and apply the code bounding theorem.

\[ H_{r}(S^{n}) \leq \overline{L}_{{\sf s}}^{\ n}(S^{n}) \leq H_{r}(S^{n})+1 \]

Substituting for the extension entropy and extension average length gives :

\[n H_{r}(S) \leq n \overline{L}_{{\sf s}}^{\ n}(S) \leq n H_{r}(S)+1 \]

Dividing  by $n$ gives:
\[ H_{r}(S) \leq \overline{L}_{{\sf s}}^{\ n}(S) \leq  H_{r}(S)+\frac{1}{n} \]

{\bf Proof Discussion}: We might want to check if the upper bound 
is attained when the entropy is zero. Only the sequence of n certain 
events in $S^{n}$ has probability one. All other sequences have 
probability zero. Since 
$\overline{L}_{{\sf s}}^{\ n}(S^{n}) =1$, we have 
$\overline{L}_{{\sf s}}^{\ n}(S) = 1/n$.  
\[ 0 \leq \frac{1}{n} \leq 0 + \frac{1}{n} \]

For an extended Shannon or better encoding of the nth extension, 
the excess of the average over the entropy approaches zero as n increases.
Another way to look at this is to realize that as we encode larger and larger 
event sets the excess bound of 1 decreases as a percent of $H_{r}$. This can 
be seen directly from the code bounding theorem. 

{\bf Theorem : Code Compression Theorem of Shannon}
There exist codes $C$ such that  
\( H_{r}(S) \leq \overline{L}_{C}(S) \leq  H_{r}(S)+\frac{1}{n} \)

Nothing is free in life. So the down side of code compression is that 
a sufficiently long message must be sent in order to apply an 
nth extension code. This means a delay in receiving messages.
In computer terms, Shannon's Noiseless Coding Theorem is a 
batch processing system. For the short messages of an interactive processing 
system, the Theorem is is not applicable. 

\section{Conclusions}
How we think about concepts is influenced by the notation used. The 
importance of notation is not a new idea [8] but does bear repeating. 
Indeed, more confusion is usually engendered by the notation used than by 
the actual ideas expressed.
 
Boundary points often exhibit unusual behaviour, and this paper gives 
an interesting example of how such behaviour is not at all obvious 
until after it has been observed.

Information theoretic inequalities are important  
statements of fundamental pattterns [7][8].
The Code Bounding and Code Compression theorems are quite elegant 
statements of the interrelationship of coding efficiency and entropy.
This new found symmetry only adds to their elegance.

\begin{acknowledgements}
Professor Paul Fieguth, Systems Design Engineering University of Waterloo, Professor Emertis Albert Stevens Univesity of New Brunswick, Ingelore Johnson\end{acknowledgements}



\end{document}